\documentclass[12pt,epsf]{article}

 \usepackage{epsfig}
 \usepackage{times}
 \usepackage{epstopdf}
 \usepackage{amsmath}
 \usepackage{amssymb}
\usepackage{color}

\newcommand{\be}{\begin{equation}}
\newcommand{\ee}{\end{equation}}
\def\({\left (}
\def\){\right )}
\def\[{\left [}
\def\[{\right ]}

\begin{document}
\begin{titlepage}
\bigskip
\rightline
\bigskip\bigskip\bigskip\bigskip
\centerline {\Large \bf {Pathologies in Asymptotically Lifshitz Spacetimes}}
\bigskip\bigskip
\bigskip\bigskip

\centerline{\large Keith Copsey$^{a,b}$ and Robert Mann$^a$}
\bigskip\bigskip
\centerline{${}^a$ \em Department of Physics and Astronomy, University of Waterloo, Waterloo, Ontario N2L 3G1, Canada}

\vspace{0.2 cm}

\centerline{${}^b$ \em Perimeter Institute for Theoretical Physics, Waterloo, Ontario N2L 2Y5, Canada}

\vspace{0.3cm}

\centerline{\em kcopsey@perimeterinstitute.ca, rbmann@sciborg.uwaterloo.ca}
\bigskip\bigskip

\begin{abstract}
There has been significant interest in the last several years in studying possible gravitational duals, known as Lifshitz spacetimes, to anisotropically scaling field theories by adding matter to distort the asymptotics of an AdS spacetime.   We point out that putative ground state for the most heavily studied example of such a spacetime, that with a flat spatial section, suffers from a naked singularity .  Furthermore, known stringy effects can not resolve this singularity without producing a regime with significant quantum corrections to the entire spacetime, including the asymptotic region. We review the reasons one might worry that asymptotically Lifshitz spacetimes are unstable and employ the initial data problem to study the stability of such systems.  Rather surprisingly  this question, and even the initial value problem itself,  for these spacetimes turns out to generically not be well-posed.  A generic normalizable state will evolve in such a way to violate Lifshitz asymptotics in finite time. Conversely, enforcing the desired asymptotics at all times puts strong restrictions not just on the metric and fields in the asymptotic region but in the deep interior as well.  Generically, even perturbations of the matter field of compact support are not compatible with the desired asymptotics.

\end{abstract}
\end{titlepage}

\baselineskip=16pt
\setcounter{equation}{0}

\section{Introduction}

Holographic concepts play an important role in theoretical physics,
demonstrating new and unexpected connections between disparate systems.
The key idea of holography is that a  non-gravitational field theory of a given dimensionality is dual
to a gravitation theory in some larger dimensionality.  The anti de Sitter - conformal field
theory (AdS/CFT) correspondence conjecture \cite{AdSCFT} is the most concrete and best explored
realization of holography. A large body of calculational evidence (although admittedly largely relying on preserving supersymmetry)
indicates that a conformal field theory (CFT)
can be mapped to string theory, including gravitational dynamics, in an asymptotically Anti de Sitter (AdS) spacetime
of one greater dimension.

However the conceptual reach of holography appears to be much wider.
It has been extended to describe a duality
between a broad range of strongly coupled field theories and gravity
in the context of QCD quark-gluon plasmas \cite{Kov}, atomic physics, and condensed matter physics \cite{Hart,Faul,Mc}. Holographic renormalization has been employed to further our
understanding of conserved quantities and gravitational thermodynamics
in both asymptotically de Sitter \cite{dSCFT} and asymptotically flat spacetimes \cite{MannMarolf}.
The full implications of  gravity-gauge duality (in, for example, elucidating the strong coupling behavior of the non-gravitational theories noted above) and the precise criterion required for systems to satisfy such a duality remain to be understood. 

In this context, a proposed duality between Lifshitz field theories and gravitation has recently received much attention.  Lifshitz field theories describe the behavior of various condensed matter systems near fixed points and exhibit anisotropic scaling relations between the temporal and spatial coordinates.  The conjectured duality involves adding a bulk coordinate $r$ to the $(t,\vec{x})$ coordinates of the Lifshitz theory such that the scaling relation
\begin{equation}
t \rightarrow \lambda^z t, \hspace{5mm} r \rightarrow \lambda^{-1} r, \hspace{5mm} \vec{x} \rightarrow \lambda \vec{x}
\end{equation}
is an asymptotic symmetry of the gravitational theory, whose natural asymptotic spacetime metric is \cite{Kachru}
\begin{equation}
ds^{2}= - \frac{r^{2z}}{l^{2z}}dt^{2}+ l^2 \frac{dr^{2}}{r^{2}}+r^{2}d\Omega%
^{2} \label{Lifmet}
\end{equation}
where $d\Omega^2$ is independent of $r$, noted earlier in a braneworld context \cite{Koroteev}.   An extended class of vacuum solutions
for a sort of higher-dimensional dilaton gravity with general $z$  appeared as an early example \cite{Tay}; shortly afterward black hole solutions, both exact (for $z=2$) \cite{Mann} and
numerical (for more general values of $z$) \cite{Mann,Peet,Bal,Dan}, were discovered.  For $z=1$ this scaling symmetry is the familiar conformal symmetry (in Poincare slicing) relevant to AdS/CFT. For $z=3$, theories with this type of scaling are
power-counting renormalizable, providing, perhaps, a UV completion to the effective gravitational field theory \cite{Horava}.  We note, however, that unlike Horava-Lifshitz approach, we will not alter the Einstein-Hilbert action to break spacetime diffeomorphism invariance at the level of the action but merely consider states that break Lorentz (and conformal) invariance and add simple types of matter to ensure we can find such solutions.  In terms of  AdS/CFT language this is simply a non-normalizable deformation, albeit one rather larger than usual. 

Perhaps the simplest such matter content consists of a massive vector field, in addition to a cosmological constant, with action for a d-dimensional spacetime
\be \label{act1}
S = \kappa \int \sqrt{-g} \Big( R - 2 \Lambda - \frac{1}{4} F_{a b} F^{a b} - \frac{m_0^2}{2} A_a A^a \Big)
\ee
where $\kappa = (16 \pi G)^{-1}$ and $F = dA$.  This action admits the exact solution for $d \geq 3$
\be \label{met1}
ds^2 = l^2 \Big(-r^{2 z} dt^2 + \frac{dr^2}{r^2} + r^2 dy^i dy_i \Big)
\ee 
\be
A = q r^z dt
\ee
where $2 \leq i \leq d-1$, provided one takes
\begin{eqnarray} \label{const1}
m_0^2 &=&  \frac{ (d - 2) z}{l^2} \qquad q^2 = \frac{2 (z - 1) l^2}{z} \nonumber \\
\Lambda &=& -\frac{z^2+(d-3) z+(d-2)^2 }{2 l^2}
\end{eqnarray}
The parametrization of the AdS-length $l$ in terms of the cosmological constant $\Lambda$ might seem awkward but it greatly simplifies the later equations and in any case is merely a convention.  
Note we have not only considered the case of (\ref{Lifmet}) where $d\Omega^2$ is flat but also rescaled the coordinates to make them dimensionless, as we will do in the remainder of this work.  Also note that reality of the fields, together with the positivity of $m_0^2$, requires $z \geq 1$ and we will restrict our attention to this case.  If one were willing to consider a tachyonic massive vector field, and the subsequent violation of the weak energy condition, one could consider $z < 0$ but this case and the resulting geometries are sufficiently exotic so we will not consider them here.   We also note this action may be taken seriously from a string perspective, since it may be obtained as a consistent truncation of ten and eleven dimensional supergravity \cite{Gauntlett08}.  

Another suitable matter content for Lifshitz spacetimes consists of a 2-form $\tilde{F}$ and a $(d-1)$-form $H$ coupled together with a Chern-Simons term with action
$$
S = \kappa \int \sqrt{-g} \Big(R - 2 \Lambda - \frac{1}{4} \tilde{F}_{a b} \tilde{F}^{a b} - \frac{1}{2 (d-1)!} H_{a_1 \ldots a_{d-1}} H^{a_1 \ldots a_{d-1}}
$$
\be \label{act2}
 - \frac{\gamma}{2 (d-2)!} \epsilon_{a_1 a_2 a_3 \ldots a_d} \tilde{F}^{a_1 a_2} B^{a_3 \ldots a_d} \Big)
\ee
where $\tilde{F} = d\tilde{A}$, $H= dB$, $\gamma$ is the Chern-Simons coupling constant, and $\epsilon_{a_1 \ldots a_d}$ is the usual volume form for a d-dimensional spacetime.  In fact, the two actions (\ref{act1}) and (\ref{act2}) are dual.  The field equations of motion from (\ref{act2}) are
\be \label{Feqn1}
\nabla_b \tilde{F}^{b a} = \frac{\gamma}{(d-1)!} \epsilon^{a a_1 \ldots a_{d-1}} H_{a_1 \ldots a_{d-1}}
\ee
and
\be \label{Feqn2}
\nabla_{a_1} H^{a_1 a_2 \ldots a_{d-1}} = \frac{\gamma}{2} \epsilon^{a b a_2 \dots a_{d-1}} \tilde{F}_{a b} = \gamma \, \epsilon^{a b a_2 \ldots a_{d-1}} \nabla_a \tilde{A}_b
\ee
In terms of forms (\ref{Feqn2}) may be written as
\be
d \star H  =  \gamma \,d\tilde{A}
\ee
Presuming our space is simply connected (and, as long as the boundary at null infinity is simply connected, topological censorship \cite{topcensor} ensures that if the spacetime is not simply connected it will be singular) there are no non-exact closed one forms so
\be
\star H = \gamma \tilde{A} - d \phi
\ee
for some scalar $\phi$ or equivalently
\be \label{Feqn3}
H^{a_1 \ldots a_{d-1}} = -\gamma \epsilon^{a a_1 \ldots a_{d-1}} \tilde{A}_a + \epsilon^{a a_1 \ldots a_{d-1}} \nabla_a \phi = -\gamma  \epsilon^{a a_1 \ldots a_{d-1}} A_a
\ee
where $A_a = \tilde{A}_a - \frac{1}{\gamma} \nabla_a \phi$.  Then, since $F = dA = d\tilde{A}$, (\ref{Feqn1}) becomes
\be
\nabla_b F^{b a} = \gamma^2 A^a
\ee
which is simply the field equation for a massive gauge field (\ref{act1}) with a mass $m_0^2 = \gamma^2$.  Further, performing an integration by parts on the Chern-Simons term in (\ref{act2}) and inserting the relationship (\ref{Feqn3}) into the result transforms the action (\ref{act2}), up to a surface term, into (\ref{act1}), completing the demonstration of the duality. 

There are, however, two major sets of concerns one might have with these spacetimes from a bulk (i.e. gravitational) point of view.  The first is an issue of regularity in the interior.  In particular, the most studied Lifshitz spacetimes have been those with what one might dub a flat section, namely (\ref{met1}).  For the case of $z =1$, the surface $r = 0$ is simply the Poincare horizon and by transforming to global coordinates one may smoothly pass though this horizon (see, e.g.,  \cite{Magoo}).  However, for $z \neq 1$ the surface $r = 0$ certainly does not appear to be a horizon.  We will describe its proper interpretation in the next section.

The second set of concerns relate to the stability of these spacetimes.  Let us first note that the usual spinorial proofs of the positive energy theorems \cite{posenergy} require an asymptotically constant spinor and hence at least asymptotic supersymmetry.  However, the asymptotics we are considering are designed to violate Lorentz invariance, and thus supersymmetry, and so they do not admit such spinors.  This alone might give one pause, since there several examples of even quite mild modifications of stable spacetimes that break all asymptotic supersymmetries and produce spacetimes with Hamiltonians which are unbounded from below \cite{Wittenbubble, CopseyMannHOP}.  Further, there is no obvious way to put the matter content above in topologically protected configurations and indeed the examples we will consider later are topologically trivial both in terms of the spacetime and matter content.  This means one is adding matter to AdS and hoping the boundary conditions alone are enough to stabilize a configuration that locally would like to collapse.  

We then study the initial data problem to address the concerns one might have about the stability of such spacetimes.   For the sake of simplicity, we restrict our attention to states with maximal transverse symmetry (i.e. take $d\Omega^2$ to be a flat, spherical, or hyperbolic metric) so that there are no independent gravitational degrees of freedom but simply those required by non-trivial matter configurations.  Rather surprisingly we find the initial data problem is not generically well-posed for these spacetimes.   The evolution of normalizable (in a Hamiltonian sense) initial data produces time dependence at the non-normalizable order (even violating the leading order asymptotics if $z \geq 2$).  Conversely, if one insists Lifshitz asymptotics are maintained at all times the solution will not be regular in the interior unless one imposes strong restrictions on the solution not just in the asymptotic region but throughout the bulk.   In particular, despite the fact that in terms of initial data one has enough freedom to specify a radial profile of the matter field arbitrarily, any perturbations that change the mass while preserving the asymptotics of the matter field are forbidden.

\setcounter{equation}{0}
\section{Flat Lifshitz solutions and interior regularity}

Let us consider issues of geodesic completeness and regularity for the exact metric often thought to be the ground state for flat $d$-dimensional Lifshitz solutions
\be \label{met2}
ds^2 = l^2 \Big(-r^{2 z} dt^2 + \frac{dr^2}{r^2} + r^2 dy^i dy_i \Big)
\ee 
where $2 \leq i \leq d-1$.  The metric (\ref{met2}) has a timelike killing vector and $d-2$ spacelike killing vectors, resulting in the conserved energies and momenta 
\be
E = -g_{t t} \dot{t}
\ee
\be
p_i = g_{i i} \dot{y}_i
\ee
and
\be
\dot{x}^{\mu} = \frac{d x^{\mu}}{d\lambda}
\ee
for some affine parameter $\lambda$. Then for a geodesic
\be
-k = g_{t t} \dot{t}^2+g_{r r} \dot{r}^2+\Sigma_i g_{y_i y_i} \dot{y}_i^2 
\ee
that is either timelike ($k = 1$) or null ($k = 0$)  we have
\be \label{geod1}
\dot{r}^2 = \frac{E^2}{l^4} r^{2 - 2 z} - \Sigma_i \frac{p_i^2}{l^4} - \frac{k r^2}{l^2}
\ee
As $r \rightarrow \infty$ the behavior of geodesics for $z > 1$ is qualitatively the same as for an asymptotically AdS space, namely that timelike geodesics never get out to infinite $r$ but null geodesics do so in finite coordinate time (with the minor caveat that if the geodesics have any momentum in the planar directions they will turn around at finite $r$).   On the other hand,  as $r \rightarrow 0$ for $z > 1$
\be
r^z \sim \pm \frac{E z}{l} (\lambda - \lambda_0)
\ee
for an appropriate constant $\lambda_0$.

Note that both timelike and null geodesics travel from finite $r$ to $r = 0$ in finite affine parameter, indicating that the space specified by (\ref{met1}) is geodesically incomplete.  Of itself this is not necessarily a disaster, for one may have merely written down coordinates covering only one part of the manifold. As previously mentioned, for $z = 1$ the above coordinates correspond to the Poincare patch with $r = 0$ the Poincare horizon. If the space is at least $C^{(1)}$, the geodesic equations of motion will be continuous and since there is enough symmetry to entirely determine the geodesics in terms of conserved quantities for the part of the manifold the coordinates cover, at least locally (near $r = 0$) one obtains another copy of (\ref{met1}) where $r(\lambda)$ is given by (\ref{geod1}).  For ingoing geodesics as $r \rightarrow 0$,
\be
r^{z} \sim \frac{E z}{l} (\lambda_0 - \lambda)
\ee
and so $r^z$ must change sign at $\lambda = \lambda_0$ .   Further, we must ensure that $r^2$ remains real and positive for $\lambda > \lambda_0$, for otherwise the metric either becomes complex or all the $g_{y_i y_i}$ become negative, as is $g_{t t}$, and the signature of the manifold would change.   Both of these conditions may be met only if $r$ extends to negative real values when $\lambda > \lambda_0$ and $z$ is an odd integer.  

The above suggests that at least for most $z$  one should expect a singularity at $r = 0$.  However in a ``static'' orthonormal basis 
\be
(e_0)_{\alpha} = -l r^z  \partial_{\alpha} t  \quad
(e_1)_{\alpha} = \frac{l}{r}  \partial_{\alpha} r  \quad (e_i)_{\alpha} = l r  \partial_{\alpha} y_i  
\ee
all the components of the affine connection and the Riemann tensor are finite and in fact constant.  Hence all curvature invariants constructed from the Riemann tensor are finite at $r = 0$. 

To address physical questions, however, we require the components of the Riemann tensor in a parallelly propagated orthonormal frame (PPON) , that is in a basis as measured by an observer traveling along a geodesic.  For our purposes it will suffice to consider radial geodesics (i.e. $p_i = 0$).  Then we want a basis with a unit timelike vector proportional to the four-velocity along such a geodesic.  Such a basis is given along a geodesic with conserved energy E by
\begin{eqnarray}
(\tilde{e}_0)_{\alpha} &=& -E  \partial_{\alpha} t \pm E r^{-1-z} \sqrt{1 - \frac{l^2 r^{2 z}}{E^2}} \,  \partial_{\alpha} r\nonumber \\
(\tilde{e}_1)_{\alpha} &=& -E  \sqrt{1 - \frac{l^2 r^{2 z}}{E^2}}  \, \partial_{\alpha} t \pm E r^{-1-z} \, \partial_{\alpha} r \\
(\tilde{e}_i)_{\alpha} &=& l r  \partial_{\alpha} y_i   \nonumber 
\end{eqnarray}
where the two choices of sign correspond to whether one is considering a radially ingoing or outgoing geodesic.  
Then, adopting the notation 
\be
R_{i j k l} \equiv R^{\alpha \kappa \gamma \delta} (\tilde{e}_i)_{\alpha}  (\tilde{e}_j)_{\kappa}  (\tilde{e}_k)_{\gamma}  (\tilde{e}_l)_{\delta}
\ee
(i.e. the components in a PPON frame) we obtain 
\begin{eqnarray}
R_{0101} &=& \frac{z^2}{l^2} \qquad 
R_{ijij} = -\frac{1}{l^2}  \, \, \, \, (i \neq j) \nonumber \\
R_{0i0i} &=& \frac{1}{l^2} + \frac{E^2 (z - 1)}{l^4 r^{2 z}} \\
R_{1i1i} &=& -\frac{z}{l^2} + \frac{E^2 (z - 1)}{l^4 r^{2 z}}  \nonumber \\
R_{0i1i} &=&  \frac{E^2 (z - 1)}{l^4 r^{2 z}} \sqrt{1-\frac{l^2 r^{2 z}}{E^2}} \nonumber
\end{eqnarray}
(where $2 \leq i,j \leq d-2$)
for the nonzero components of the Riemann tensor.
Hence tidal forces diverge as $r \rightarrow 0$ if $z\neq 1$.

\begin{figure}
\begin{picture} (0,0)
         \put(60, -110){$r = \infty$}
    \end{picture}
    \centering
    
\includegraphics[scale= 0.75]{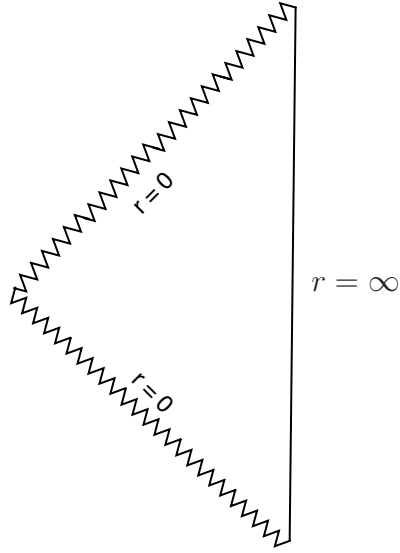}
	\caption{Penrose diagram for flat Lifshitz spacetimes}
	\label{ex1fig}
	\end{figure}

Note that the normal to surfaces of constant $r$ is spacelike at any nonzero $r$ but beomes null as $r \rightarrow 0$, just as in the Poincare slicing of AdS
\be
\nabla_a r \nabla^a r = g^{r r} = \frac{r^2}{l^2}
\ee
so $r = 0$ is a null surface.  This can also be seen in terms of the metric by defining
\be
\tau  = t - \frac{1}{z r^z}
\ee
and
\be
u = \frac{2}{z} r^z
\ee
since then (\ref{met2}) becomes
\be
ds^2 = l^2 \Big[ - \frac{z^2}{4} u^2 d\tau^2 + d\tau du + \Big(\frac{z u}{2} \Big)^{2/z} dy^i dy_i \Big]
\ee
Then, given the above, the spacetime has a null curvature singularity similar that of singular gravitational plane waves \cite{Horosteif}.   We emphasize that the above observation regarding  large tidal forces near $r = 0$ has been noted before \cite{Kachru},  as has a description of the $r = 0$ surface as a null singularity along with some details of the divergence for null geodesics \cite{Hart}.  Since the spacetime possesses time symmetry and has both ingoing and outgoing null rays, the Penrose diagram (see Figure \ref{ex1fig}) for the Lifshitz spacetime (\ref{Lifmet}) looks like the Poincare patch, except with singularities along what would be a Poincare horizon for $z=1$.   Since any observer can see the past null singularity the metric (\ref{Lifmet})  describes a spacetime with a naked singularity.   

Note the fact that $r=0$ is a null surface means the ``static'' orthonormal frame is not one any physical observer can reach with a finite boost.  The frame is, however, useful in allowing us to constrain $\alpha'$ effects.   The components of the field strength and Riemann tensor (as well as the connection) contracted into the static frame one-forms, are finite near $r=0$ and in fact constant throughout the space.   Hence at least for sufficiently small $l_p/l$, $\alpha'$ corrections will be approximately constant throughout the space and thus can not become large near $r = 0$ (affording at least the possibility of resolving the singularity) without becoming significant throughout the space, including in the asymptotic region.  As one increases $l_p/l$ non-linear effects among the Planck-suppresed terms become significant, although it is difficult to imagine a scenario where the asymptotic region is dominated by stringy effects for very small $l_p/l$ but becomes approximately classical, at least for some range, as one increases $l_p/l$. Furthermore  these solutions have not required a non-trivial dilaton and  the consistent embeddings of Lifshitz spacetimes into string theory that we are aware of do not involve the dilaton becoming large at $r =0$ while remaining small asymptotically.  In particular, explicit lifts involving a constant dilaton \cite{Gauntlett08} and  a dilaton that is independent of the Lifshitz directions \cite{Gregory2010} have both been presented.  Hence it appears futile to appeal to string loop effects to resolve this singularity.\footnote{This feature distinguishes the present case from extremal Dp-branes for $p > 3$ where one again has a null singularity in the metric \cite{HorowitzRoss} but one also has a diverging dilaton and so stringy corrections can resolve the singularity.}  Note the proposition that one loses control of the calculation throughout a low-curvature, weakly coupled region, especially considering the embeddings of these solutions and action into honest supergravity solutions, would actually be quite radical.  Hence the conservative perspective would seem to be to conclude  these singularities should be regarded as pathological in string theory, as well as classically.  

One might be concerned that the above argument contradicts the conjecture of Gubser \cite{Gubser:2000nd} connecting acceptable singularities to those that may be cloaked by an event horizon.   To the best of our knowledge, there is in fact no contradiction at present between this conjecture and the above argument.  While one may construct static Lifshitz black holes with regular horizons, as we have noted above, the solutions constructed to date describe eternal black holes and, as usual, have a naked singularity in the past.   One might presume that, as in flat space, one could begin with a regular spacetime and collapse matter to form a black hole without this undesirable feature.   However, to carry out this program we would need an asymptotic  flat  Lifshitz spacetime that is free of (naked) singularities in the interior to begin with.  As we will now discuss, at least the obvious candidates for such solutions either fail to exist or have hidden singularities. 

We then wish to search for states with Lifshitz asymptotics that are also sensible in the interior, both to try to find some regular (or at least not nakedly singular) solution that corresponds to the ground state for Lifshitz asymptotics and to explore the possible tension with Gubser's conjecture.  Let us consider all static solutions which, like (\ref{met2}), might be described as plane symmetric and has only an electric field\footnote{The absence of a radial magnetic field is actually no restriction since this would result in time dependence, as we will later see explicitly.  This result should not be surprising since a radial magnetic field would result in electromagnetic momentum.}
\begin{eqnarray}
ds^2 &=& l^2\Big( f_1(r) dt^2 + \frac{dr^2}{W(r)} + r^2 dy_i dy^i \Big) \nonumber \\
A &=& \alpha(r) dt
\end{eqnarray}
and asymptotically Lifshitz, namely that asymptotically $f_1(r) \rightarrow -r^{2 z}$ and $W(r) \rightarrow r^2$.  The Einstein and field equations are equivalent to
\be \label{stateqn1}
f_1'(r)+\frac{f_1(r)}{r} \Big(d-3 + \frac{2 l^2 \Lambda r^2 }{(d-2) W(r)}  \Big)  =  \frac{r}{2 (d-2)} \Big(\frac{(\alpha'(r))^2}{l^2} - \frac{ m_0^2 \alpha^2(r)}{W(r)} \Big)
\ee
\be \label{stateqn2}
W'(r) +(d-3) \frac{W(r)}{r} + \frac{2 l^2 \Lambda r}{d-2} =\frac{r}{2 (d-2) f_1(r)} \Big( m_0^2 \alpha^2(r) + \frac{W(r) (\alpha'(r))^2}{l^2} \Big)
\ee
and
\be \label{stateqn3}
\alpha''(r) = -(d-2) \frac{ \alpha'(r)}{r} - \frac{m_0^2 \, r \alpha^2(r) \alpha'(r)}{2 (d-2) f_1(r) W(r)} + \frac{ l^2 m_0^2 \alpha(r)}{W(r)}
\ee
Let us first consider whether there are any possible values of the constants $m_0$ and $\Lambda$ or  asymptotics for $\alpha(r)$, besides those in our previous exact solution (\ref{const1}), consistent with the desired metric asymptotics.   Solving algebraically (\ref{stateqn1}) and (\ref{stateqn2}) for $\alpha(r)$ asymptotically, one finds $\alpha^2(r) \rightarrow {2 (d-2)  (z - 1) r^{2 z}/{m_0^2}}$ as $r \rightarrow \infty$.   Using the remaining equations at leading order, one finds $m_0$ and $\Lambda$ are required to take precisely the same values as in (\ref{const1}).  In other words, the asymptotics for the potential and constants $m_0$ and $\Lambda$ we had previously considered are not just sufficient but necessary conditions to have a metric with the given symmetries and desired asymptotics.

In searching for a geometry for a flat-section asymptotically Lifshitz solution that corresponds to a ground state and is regular in the deep interior,  the two obvious possibilities are a usual Poincare-type horizon at $r = 0$ or an extremal black hole.\footnote{The matter content here respects the null energy condition so via the usual argument with the Raychaudhuri equation one expects wormhole-type solutions are impossible.  Concretely, if one tries to find a solution where $W(r) \rightarrow (r-r_0)^2$ and $f_1(r_0) \neq 0$, (\ref{stateqn2}) shows $\alpha(r_0) \neq 0$ (and $f_1(r_0) < 0$) and this is not compatible with (\ref{stateqn1}).}   In the case of the first possibility, we must assume that as $r \rightarrow 0$, $W(r) \approx w_0 r^2$ and $f_1(r) \approx -w_0 r^2$ for some constant $w_0$.  Note of course the constant for $f_1(r)$ is arbitrary as far as any local considerations are concerned; one could always rescale $t$ to set it to any desired value.  Now examining (\ref{stateqn2}), this means the left hand side has a determined limit as $r \rightarrow 0$.  Since both terms on the right hand side of (\ref{stateqn2}) have the same sign, then as $r \rightarrow 0$ either $\alpha(r) \rightarrow 0$ faster than r  (and the left hand side of (\ref{stateqn2}) vanishes at leading order) yielding
\be \label{wodef}
w_0 = -\frac{2 \Lambda}{(d-1) (d-2)}
\ee
or otherwise
\be \label{acond1}
\alpha(r) \rightarrow a_0 r
\ee
for
\be \label{a0def}
a_0^2 = -\frac{2 l^2 w_0 (2 l^2 \Lambda + (d-1) (d-2) w_0)}{w_0 + l^2 m_0^2}
\ee
However  inserting (\ref{a0def}) into (\ref{stateqn1}) at leading order one quickly finds that (\ref{wodef}) is required anyway and $a_0 = 0$.   Then we must take (\ref{wodef}) and $\alpha(r)$ going to zero faster than $r$ as $r \rightarrow 0$.  Solving (\ref{stateqn1})-(\ref{stateqn3}) at leading order as $r \rightarrow 0$ one finds
\be
\alpha(r) \rightarrow a_1 r^{n_1}
\ee
where
\be
n_1= -\frac{d-3}{2} +\frac{1}{2} \sqrt{ (d-3)^2 + \frac{4 z (d - 1) (d-2)^2}{z^2 + (d-3) z + (d-2)^2}}
\ee
Plotting $n_1$ for $d = 4$ one finds a range of $z$ for which $n_1 > 1$. However $n_1$ is much smaller than one would expect; in fact for $d = 4$ it is maximized at $z= 2$ where $n_1 \approx 1.1279$.   These small fractional powers suggest one should be concerned about the regularity of the solution as $r \rightarrow 0$, for these powers will feed into the metric functions via (\ref{stateqn1}) and (\ref{stateqn2}) and since the Riemann tensor involves two derivatives certain components of the Riemann tensor threaten to diverge. 

If one takes the obvious static orthonormal basis
\be
(e_0)_{\alpha} =   - l \sqrt{-f_1(r)} \partial_{\alpha} t  \quad
(e_1)_{\alpha} = l \frac{1}{\sqrt{W(r)}}  \partial_{\alpha} r  \quad (e_i)_{\alpha} = l r  \partial_{\alpha} y_i  
\ee
all the components of the affine connection and the Riemann tensor, as well as the field strength, contracted into this basis are finite at $r = 0$.   Hence all curvature invariants from the Riemann tensor are finite as well.  

 As above, however, this does not guarantee the absence of singularities and hence we again consider the components of the Riemann tensor in a parallelly propagated orthornormal frame.  Again limiting our attention to considering radial geodesics and finding a basis with a unit timelike vector prportional to the four-velocity along radial geodesics with conserved energy E
\begin{eqnarray}
(\tilde{e}_0)_{\alpha} &=& -E  \, \partial_{\alpha} t \pm l \sqrt{\frac{-1 - \frac{E^2}{l^2 f_1(r)}}{W(r)}}\,  \partial_{\alpha} r\nonumber \\
(\tilde{e}_1)_{\alpha} &=& - \sqrt{E^2 +  l^2 f_1(r)} \, \partial_{\alpha} t \pm \frac{E}{\sqrt{-f_1(r) W(r)}}  \, \partial_{\alpha} r \\
(\tilde{e}_i)_{\alpha} &=& l r  \partial_{\alpha} y_i   \nonumber 
\end{eqnarray}
where the upper and lower signs correspon to radially outgoing and ingoing geodesics respectively.  Using the equations of motion (\ref{stateqn1})-(\ref{stateqn3}) one can show unless $\alpha(r) \sim r^2$ or faster as $r \rightarrow 0$  components of the Riemann tensor contracted with this PPON diverge
\be
R_{a i b  i } \rightarrow \frac{ (d-1)^2 (d-2)^2 z E^2}{8 l^{10} \Lambda^2} \frac{\alpha^2(r)}{r^4}
\ee
where $a$ and $b$ are either $0$ or $1$.  Examining $n_1$ it is easy to check there is a range of $z$ where $n_1 \geq  2$ only if $d \geq 8$.   Hence, at least up through seven dimensions, the naively smooth solution which tries to interpolate between Lifshitz asymptotics, including the renormalization group flow solution of \cite{Kachru}, has a naked singularity very much like the naive ground state discussed before.  Although for higher dimensions there is no such obvious problem,  we are not aware  of either explicit constructions of such solutions or more importantly embeddings of such high dimensional Lifshitz solutions into ten or eleven dimensional supergravity solutions.  

For our last attempt in finding a reasonable ground state for these asymptotics, let us suppose  the above equations admit extremal black holes.  Then there must be some constant $r_0$ such that as $r \rightarrow r_0$, $W(r) \rightarrow w_1 (r-r_0)^2$ and $f_1(r) \rightarrow -t_1 (r - r_0)^2$ for some positive constants $t_1$ and $w_1$.  Proceeding as before, (\ref{stateqn2}) then requires as $r \rightarrow r_0$,
\be \label{alph2}
\alpha(r) \rightarrow a_2 (r - r_0)
\ee
where
\be
a_2^2 =\frac{2 ( z^2 + (d-3) z + (d-2)^2)}{ (d-2) z+w_1} l^2 t_1 = -\frac{4 l^4 \Lambda t_1}{(d-2) z+w_1}
\ee
Then (\ref{stateqn3}) would imply that as $r \rightarrow r_0$,
\be
\alpha''(r) \rightarrow \frac{z r_0}{2 l^2} \frac{a_2^3}{w_1 t_1 (r-r_0)^2}
\ee
implying that $\alpha(r)$ is logarithmically divergent as $r \rightarrow r_0$, in contradiction with (\ref{alph2}).  While a very broad-minded reader might wonder if the above is overly restrictive in assuming $w_1>0$, the above is still correct at leading order unless one tunes $w_1 < 0$ such that (\ref{stateqn2}) at leading order is consistent with $\alpha \sim (r-r_0)^{n_0}$ where $n_0 < 1$.   It is straightforward to check this scenario with a diverging field strength is not compatible with (\ref{stateqn3}).

Finally, one might simply try to content oneself with studying black holes with these asymptotics.  While indeed the solutions of this type appear to be perfectly acceptable, note the existence of a regular black hole (i.e. one with a regular horizon) does not mean the theory one is dealing with is sensible and in particular has a well-defined ground state.  Probably the most familiar example is found in Kaluza-Klein theory by taking boundary conditions of antiperiodic fermions around the asymptotic Kaluza-Klein directions; one may construct perfectly regular black holes in this theory, despite the fact such boundary solutions admit regular states of arbitrarily negative energy \cite{Wittenbubble}.  Another, even more extreme, example is the fact that one could collapse a positive mass shell around a negative mass Schwarzschild black hole to produce a perfectly regular positive mass black hole. However, if the singularity of negative mass Schwarzschild were to be resolved by quantum gravity effects it would be disastrous for string theory or any other theory of quantum gravity \cite{HorowitzMyers}.

\setcounter{equation}{0}
\section{Lifshitz spacetimes and the initial value problem}

\subsection{Hamiltonian formalism}

As mentioned previously, one might worry that the matter content supporting the Lifshitz asymptotics could relax away.  For fixed boundary conditions this would, at least generically, translate into the statement that the spacetime with the given asymptotics admit states of arbitrarily negative energy.  We need not assume such states are stationary (indeed one would be surprised if they were) or construct a full spacetime solution but merely construct states consistent with Einstein's equations with some fixed conserved energy.  Generically, of course, the states will time evolve, in a manner specified by the Einstein and field equations, but this evolution will not change the conserved energy.  In the language of Hamiltonian mechanics, we need only satisfy the initial data constraints.  The value of the energy for these spacetimes may be directly obtained by finding the on-shell value of the Hamiltonian for these states.  Previous definitions of the energy have been given in terms of holographic renormalization \cite{RossHologLifshitz,Deh3}, in the case of a flat spatial boundary metric, and background subtraction \cite{Peet}.  Presumably, all these definitions agree up to zero point ambiguities and subtleties involving the definition of ``normalizable'' modes within the respective approaches\footnote{Indeed, a disagreement between between the background subtraction and holographic renormalization definitions of normalizability is known to take place for $z \leqq 2$.}, although we will not seek to make that comparison here.   Rather we simply take the perspective that any sensible definition of the energy must be equivalent, up to zero point ambiguities, to the on-shell value of the Hamiltonian.

For the sake of simplicity we will restrict our attention to four dimensions and consider the action for a massive vector field,
\be
S = \kappa \int \sqrt{-g} \Big( R - 2 \Lambda - \frac{1}{4} F_{a b} F^{a b} - \frac{m_0^2}{2} A_a A^a \Big)
\ee
where $\kappa = (16 \pi G)^{-1}$ and $F = dA$ and 
\be
m_0^2 = \frac{2 z}{l^2}
\qquad  
\Lambda = -\frac{(z^2+z+4)}{2 l^2}
\ee
We will consider spacetimes with Lifshitz asymptotics, namely that asymptotically
\begin{eqnarray}
ds^2 &\rightarrow& l^2\Big(-r^{2 z} dt^2 + \frac{dr^2}{r^2} + r^2 d\Omega^2\Big) \nonumber \\
A &\rightarrow& q r^z dt
\end{eqnarray}
where
\be
q^2 = \frac{2 (z - 1) l^2}{z}
\ee
and $d \Omega^2$ is not required to be flat, although for technical simplicity we will later require it to be a constant curvature space (i.e. plane, sphere, or hyperboloid).  As we will see below and has been observed in the construction of various explicit solutions \cite{Mann,Peet,Bal,Dan}, the curvature of $d\Omega^2$, if any, will enter asymptotically only at subleading order and so we take $\Lambda$ and $m_0$, as well as the asymptotics of $A$ to be consistent with their values in the flat case (\ref{met1}, \ref{const1}).  

We then perform the usual Hamiltonian decomposition into space and time for a spacelike slice $\Sigma$ with unit timelike normal $n^a$. The spatial metric induced on the surface is given by $h_{a b} = g_{a b} + n_a n_b$.  It will be useful to define the potential as
\be
\phi = n^a A_a
\ee
and the ``electric field''
\be
E^a = n_b F^{b a}
\ee
The canonical momentum differs from the above ``electric field'' by a factor of the determinant $h$ of the metric  $h_{a b}$
\be
\pi^a = \frac{\partial \mathcal{L}}{\partial \dot{A}_a} = \kappa \sqrt{h} E^a
\ee
where the time derivative of $A_a$ is given by the projection of the Lie derivative into the surface $\Sigma$
\be
\dot{A}_a = h_a^b \mathcal{L}_\xi A_b
\ee
and the time evolution vector $\xi$ may be decomposed as usual into the lapse and shift as
\be
\xi^a = N n^a + N^a
\ee
where
\be
N = -n_a \xi^a
\ee
and
\be
N^a = h^{a}_b \xi^b
\ee
The canonical Hamiltonian density derived from the above Lagrangian density $\mathcal{L}$ is
\be
\mathcal{H} = \pi^a \dot{A}_a + {\pi^{(G)}}^{a b} \dot{h}_{a b} - \mathcal{L}
\ee
where the momentum canonically conjugate to the spatial metric ${h}_{ab}$
is, as usual,
\begin{equation}\label{piG}
\pi_{G}^{ab} = \frac{\partial \mathcal{L}}{\partial
\dot{{h}}_{ab}} = \kappa \sqrt{{h}}
({K}^{ab} - {h}^{ab} {K} )
\end{equation}
where ${K}^{ab}$ is the extrinsic curvature, ${K}=
{K}^{ab}{h}_{ab}$, and
\be
\dot{h}_{a b} = h_a^c h_b^d \mathcal{L}_\xi h_{c d}
\ee
Up to surface terms, which we will deal with shortly, the canonical Hamiltonian is the generator of time translations and thus vanishes on-shell.  In other words, the Hamiltonian takes the pure constraint form
\be \label{Hform}
H = \int N C_0 + N^a C_a + \xi^a A_a \, \mathcal{C}
\ee
Each of the above $C$'s corresponds to a constraint--that is one of Einstein's equations or a field equation with no second time derivatives and hence which must be satisfied by any initial data.  The remaining Einstein and field equations then specify the time evolution of that state.   
The scalar constraint is
\begin{eqnarray} \label{scalconst}
C_0 &=& 2 \kappa \sqrt{h} (T_{a b} - G_{a b}) n^a n^b \nonumber\\
&=&
-\kappa \sqrt{h} R^{(3)}+\frac{1}{\kappa \sqrt{h}}\Big({\pi^{(G)}}^{a b} \pi^{(G)}_{a b} -\frac{{\pi_{(G)}^2}}{2}\Big) + 2 \kappa \sqrt{h} \Lambda
\nonumber \\
&&\quad +\kappa \sqrt{h} \frac{m_0^2}{2} (\bar{A}^a \bar{A}_a+\phi^2)+ \frac{\kappa \sqrt{h}}{4} \bar{F}_{a b} \bar{F}^{a b} + \frac{\pi^a \pi_a}{2 \kappa \sqrt{h}}
\end{eqnarray}
where the spatial component of the potential and field strength are given by
\begin{eqnarray}
\bar{A}_a &=& h_a^b A_b \nonumber \\
\bar{F}_{a b} &=& h_a^c h_b^d F_{c d}
\end{eqnarray}
and $R^{(3)}$ is the Ricci scalar calculated using the spatial metric $h_{a b}$.  
The momentum constraint is
\begin{eqnarray} \label{momconst}
C_a &=& 2 \kappa \sqrt{h}(T_{c b} - G_{c b}) h_a^c n^b
\nonumber\\
&=&-2 \sqrt{h} D_b\Big(\frac{{\pi^{(G)}}^b_a}{\sqrt{h}}\Big) + \bar{F}_{a b} \pi^b + m_0^2 \kappa \sqrt{h} \phi \bar{A}_a
\end{eqnarray}
where $D_a$ is the covariant derivative compatible with $h_{a b}$.  Finally, the gauge constraint is
\be
\mathcal{C} = -\kappa \sqrt{h}\Big[ D_a \Big(\frac{\pi^a}{\kappa \sqrt{h}} \Big) +  m_0^2 \phi \Big]
\ee

The appropriate surface terms for the Hamiltonian are determined by demanding that when one performs integration by parts on the above pure constraint Hamiltonian to derive the equations of motion, the resulting surface terms cancel with those we add by hand.  If one fails to do this, the desired bulk equations of motion do not actually extremize the Hamiltonian, or in other words the Hamiltonian does not have a good variational principle.  To ensure that it does,  we must add the surface terms
\begin{eqnarray} \label{Hsurf}
H_s &=& \kappa \int dS^a h^{b c} \Big[N D_c(\delta h_{a b})-N D_a (\delta h_{b c}) - D_c(N) \delta h_{a b} + D_a (N) \delta h_{b c} \Big]
\nonumber \\
&&\quad +\int dS_a \Big[ 2 N_b \frac{\delta \pi^{a b}}{\sqrt{h}}+2 N^c  \frac{\pi^{a b}}{\sqrt{h}} \delta h_{b c} - N^a  \frac{\pi^{b c}}{\sqrt{h}} \delta h_{b c}\Big]
\nonumber \\
&&\quad
-\int dS_a N \kappa \sqrt{h} \bar{F}^{a b} \delta \bar{A}_b - \int dS_a N^a \pi^b \delta \bar{A}_b + \int dS_b \pi^b N^a \delta \bar{A}_a
\nonumber \\
&&\quad
+\int dS_a \xi^b A_b \frac{\delta \pi^a}{\sqrt{h}}
\end{eqnarray}

\subsection{Defining the energy}

Recall we wish to define the energy via the on-shell value of the Hamiltonian, which implies the value of the energy is simply given by evaluation the above surface terms (\ref{Hsurf}), since the constraint terms vanish on-shell.  For the sake of simplicity, we confine our attention to  the simplest initial data generalizing the previous exact flat solution (\ref{met1}), that is
 a metric of the form
\be \label{met3}
ds^2 = l^2 \Big[\frac{dr^2}{W(r)} + r^2 k_{i j}(y) dy^i dy^j \Big]
\ee
along with
\be
\bar{A}_a = 0
\qquad 
{\pi^{(G)}}^{a b} = 0
\ee
where the only nonzero components of the fields are $\phi(r)$ and $\pi^{r}(r)$ (as well as, of course, cosmological constant) consistent with the statement that as $r \rightarrow \infty$
\be
A_t \rightarrow q r^z
\ee
To match Lifshitz asymptotics, we require that asymptotically
\begin{eqnarray}
W(r) &\rightarrow& r^2 \nonumber \\
N &\rightarrow& r^z \nonumber \\
N^a &\rightarrow& 0
\end{eqnarray}
Given $\bar{A}_a = 0$ and ${\pi^{(G)}}^{a b} = 0$, a nonzero $N^a$ will not enter into our analysis and we need not concern ourselves with the rate at which it must falloff (or generalizing the discussion to the case where it does not falloff asymptotically).  We also assume $k_{i j}$ describes a constant curvature space (if one does not do so, generically $g_{r r} (r, y_i)$ and the solution of the constraint is difficult to obtain explicitly) with any given sign of the curvature, that is $R^2$, $S_2$, or the hyperbolic metric $H_2$.  In the last case we note one is free to make identifications on the hyperbolic space to make it compact \cite{topBH}.
While we use the above as technically simplifying assumptions, it is reasonably clear from the constraint (\ref{scalconst})  that any other contributions will only serve to increase the energy. In particular, there does not seem to be any reason to believe that allowing nonzero gravitational momentum or a generic $k_{i j}(r,y)$ produces any different effects from the usual $z = 1$ AdS case where one has the standard positive energy theorems \cite{posenergy}.

We now wish to consider the boundary conditions we must impose of the fields such that the Hamiltonian will be finite, or, in language more familiar to string theorists, that the perturbations will be normalizable.  At  this point it becomes convenient to factor out the asymptotic behavior $\pi$ and introduce a new function $f(r)$ such that
\be \label{piexp}
\frac{\pi^r}{\kappa} =  -q z r^2 \Big[1+\frac{f(r)}{r^2}\Big] \sqrt{k^{(0)}}
\ee
where $k^{(0)}$ is the determinant of $k_{i j}$.   The standard Heannaux-Teitelboim  \cite{Henneaux:1984xu} type boundary conditions require that each of the individual terms in the above Hamiltonian is finite or equivalently that the normalizable part of the field falls off at least as fast as
\be \label{pifalloff1}
\delta f \sim r^{-z}
\ee
and the normalizable part of $W(r)$ (i.e. the normalizable metric perturbation) falls off at least as fast as
\be\label{Wasym}
\delta W \sim r^{-z}
\ee
 One may potentially still ensure a finite Hamiltonian for slower falloff rates provided one agrees to impose, as a boundary condition, extra correlations between the asymptotic metric and asymptotic vector field.   As a matter of principle the status of such boundary conditions is not entirely clear, but even if one agrees to allow them this will not cure the problems we will discuss below.
 
Then, presuming one takes the above standard boundary conditions, we may separate out the normalizable and non-normalizable parts of the vector field as
\be
f(r) = f_0(r) + \gamma(r) r^{-z}
\ee
where $f_0(r)$ is the non-normalizable asymptotic part of the field specified as part of the boundary conditions and asymptotically $\gamma(r)$ approaches some finite (possibly zero) value in accordance with (\ref{pifalloff1}).  Likewise for the metric 
\be
W(r) = W_0 (r) - \mu(r) r^{-z}
\ee
where $W_0(r)$ is regarded as fixed by the boundary conditions ($W_0(r) = r^2 + \ldots$) and  $\mu(r)$ (often known in other contexts as the ``mass function'') corresponds to the normalizable piece and (by (\ref{Wasym})) approaches a constant asymptotically.    
Given all of the above, the energy becomes
\be \label{Enice}
E = \kappa \Omega_k 2 l^2 \Big( \mu(\infty) - (z - 1) \gamma(\infty) \Big)
\ee

One might be surprised that the above expression for the energy contains a contribution not just from the asymptotic part of the metric but from the asymptotic electric field as well.  The latter  contribution arises from the last surface term in (\ref{Hsurf})
\be
\int dS_a \xi^b A_b \frac{\delta \pi^a}{\sqrt{h}}
\ee
In asymptotically AdS (or asymptotically flat) spacetimes $\xi^a A_a$ approaches a constant and one has enough gauge freedom to insist that $\xi^a A_a$ vanishes asymptotically.  Indeed, in that situation, the value of the Hamiltonian for a charged system is not well defined until the asymptotic value of $\xi^a A_a$ is specified.  In simple language the energy is not fixed for standard AdS charged solutions until the potential at infinity is fixed.  

Here we have a rather more delicate situation and if one tried to define the energy without considering such a term the resulting expression would not be diffeomorphism invariant.   Consider the asymptotic radial redefinition
\be \label{radialdiff}
r = \bar{r}(1 + \alpha \, \bar{r}^{-z-2})
\ee
for some constant $\alpha$ and where, if it is not immediately apparent, the power is chosen such that the above falloff conditions for the metric are preserved (i.e. the gravitational surface terms in the Hamiltonian are finite and we have a proper diffeomorphism).   Then, consider the generalization of the above (\ref{met3}, \ref{piexp}) where $r$ is not gauge fixed beyond the normalizable order (i.e. one allows various definitions of $r$ consistent with (\ref{radialdiff}))
\begin{eqnarray} \label{genasym}
ds^2 &=& \frac{dr^2}{W_0(r) - \frac{\mu(r)}{r^z}} + r^2 \Big(k^{(0)}_{i j}(y) + \frac{\delta k_{i j}(r,y)}{r^{z+2}}\Big) dy^i dy^j \nonumber \\
\frac{\pi^r}{\kappa} &=&  -q z \Big[r^2 +f_0(r) + \gamma(r) r^{-z} \Big] \sqrt{k^{(0)}} 
\end{eqnarray}
and we assume $\delta k_{i j}$ asymptotically has a finite value.  
Then a similiar calculation to that above shows that the gravitational terms yield
\begin{eqnarray}
E_g &\equiv& \kappa \int dS^a h^{b c} \Big[N D_c(\delta h_{a b})-N D_a (\delta h_{b c}) - D_c(N) \delta h_{a b} + D_a (N) \delta h_{b c} \Big] \nonumber \\
 &=& \kappa \int d\Omega \sqrt{k^{(0)}}  l^2 \Big[ 2 \mu(\infty) +  (2 z+1) \delta k(\infty) \Big]
 \end{eqnarray}
 where $\delta k (\infty) = \lim_{r \rightarrow \infty} {k^{(0)}}^{i j} \delta k_{i j}(r,y) $.
A few lines of algebra shows that the expressions for $\mu(\infty)$ and $\delta k_{i j}(\infty)$ are related to the corresponding quantities calculated in the $\bar{r}$ coordinate system by
\begin{eqnarray}
\bar{\mu}(\infty) = \mu(\infty) - 2 (z + 2) \alpha \nonumber \\
\delta \bar{k}_{i j} (\infty) = \delta k_{i j} (\infty) + 2 \alpha {k^{(0)}}_{i j}
\end{eqnarray}
and hence
\be
2 \bar{\mu} (\infty) + (2 z+1) \delta \bar{k} (\infty) = 2 \mu(\infty) + (2 z+1) \delta k (\infty) + 4 (z - 1) \alpha
\ee
Hence for $z\neq 1$ this definition of energy would not be diffeomorphism invariant.  However, noting (\ref{genasym}), under this diffeormorphism $\gamma(\infty)$ shifts by a constant
\be \label{gammashift}
\bar{\gamma}(\infty) = \gamma(\infty) + 2 \alpha
\ee
Note then any attempt to regard $\gamma(\infty)$ as a boundary condition is not consistent with this diffeomorphism. Perhaps more importantly, the shift (\ref{gammashift}) is precisely what is required to make the value of the Hamiltonian invariant under the diffeormorphism (\ref{radialdiff}).

\subsection{Solving the constraints}

Given the above symmetry and matter field assumptions, the scalar constraint becomes
\be \label{simpscalconst}
R^{(3)} = 2 \Lambda + \frac{m_0^2}{2} \phi^2 + \frac{\pi^a \pi_a}{2 \kappa^2 h}
\ee
the momentum constraint becomes trivial, and the gauge constraint becomes
\be
D_a\Big(\frac{\pi^a}{\kappa \sqrt{h}} \Big) = \frac{1}{\sqrt{h}} \partial_a\Big(\frac{\pi^a}{\kappa}\Big) = -m_0^2 \phi
\ee
Recalling the previous definition for $f(r)$ (\ref{piexp})
\be
\frac{\pi^r}{\kappa} =  -q z r^2 \Big[1+\frac{f(r)}{r^2}\Big] \sqrt{k^{(0)}}
\ee
the gauge constraint is equivalent to the statement that
\be \label{phiconst}
\phi = q l^2 r \frac{\sqrt{k^{(0)}}}{\sqrt{h}} \Big[1+\frac{f'(r)}{2 r} \Big]
\ee
Likewise, the scalar constraint (\ref{simpscalconst}) becomes 
\be \label{sconst3}
R^{(3)} = 2 \Lambda + \frac{(z-1) W(r)}{2 l^2 r^4} (2 r + f'(r))^2+\frac{z (z-1)}{r^4 l^2} (f(r)+r^2)^2
\ee
or
\be \label{constsol}
W'(r) + b_0(r) W(r) = b_1(r)
\ee
where
\begin{eqnarray}
b_0 &=& \frac{1}{r} + \frac{(z-1)}{4 r^3} (2 r +f'(r))^2 \nonumber \\
b_1 &=& \frac{R^{(0)}}{2 r} - \Lambda l^2 r - \frac{z (z-1)}{2 r^3} \Big( r^2+ f(r) \Big)^2
\end{eqnarray}
and $R^{(0)}$ is the Ricci scalar calculated using $k_{i j}(y)$.

Note then while the constraints determine $\phi$ exactly (\ref{phiconst}) and the metric in terms of a first order ordinarily differential equation (\ref{constsol}), we are still free to specify $f(r)$ to be an arbitrary function.   To deal with the remaining freedom in (\ref{constsol}), note that provided that $f(r)$ eventually falls off (to be precise, asymptotically $f'(r) \ll r$), $b_0(r) \rightarrow z/r$ and (\ref{constsol}) does not fix a term in $W(r)$ that asymptotically goes as $r^{-z}$.   Once one insists that the solution is regular--namely that $W(r)$ does not diverge at the origin or vanishes at the horizon, depending on the case under consideration--this freedom will be fixed.  Physically this contribution will enter into the mass ($\mu(\infty)$ in (\ref{Enice}) if one takes conventional boundary conditions) and reflects the fact that the energy depends on the behavior of $f(r)$ throughout the spacetime.

\subsection{Exact solutions}

Let us pause for a moment to consider exact solutions.  It is rather difficult to find $f(r)$ such that one can explicitly integrate (\ref{constsol}) but there is one exception--namely when $f(r)$ is a constant.  In the case $R^{(0)} = 0$ the only such static solutions are simply the previously known exact flat solution (\ref{met1}).  For $R^{(0)} \neq 0$ one recovers two additional classes of exact solutions.  In the first, $z=2$ and $f(r)= 0$ with the result
\begin{eqnarray} \label{exactz2sol}
ds^2 &=& l^2\Big[-r^2 \Big(r^2 + \frac{k}{2} \Big) dt^2 +\frac{dr^2}{r^2 + \frac{k}{2} } + r^2 d\Omega^2 \Big] \nonumber \\
A_t &=& l \Big( r^2 + \frac{k}{2} \Big)
\end{eqnarray}
where $d\Omega^2$ is a unit sphere or hyperbolic metric, depending on whether $k = \pm 1$
\be \label{dOmega}
d\Omega^2 = d\theta^2 + \frac{\sin^2(\sqrt{k} \theta)}{k} d\phi^2
\ee
These solutions, first found by numerically inspired guesswork \cite{Mann}, describe a naked singularity (at $r = 0$) if $k = 1$ and a black hole with a regular horizon if $k = -1$.

The second class of static solutions, also previously found by guesswork in the case of positive $R^{(0)}$ \cite{Peet}, exist provided $z =4$ and describe black holes with regular horizons for $k = \pm 1$
\begin{eqnarray}
ds^2 &=&  l^2\Big[-r^6 h(r) dt^2 +\frac{dr^2}{h(r)} + r^2 d\Omega^2 \Big] \nonumber\\
A_t &=& l \sqrt{\frac{3}{2}} r^2 h(r)
\end{eqnarray}
where
\be
h(r) = r^2 + \frac{k}{10} -\frac{3}{400 r^2}
\ee
and $d \Omega^2$ is as before (\ref{dOmega}).

\subsection{Time evolution of initial data}

As we have remarked before, we may solve the constraints with an arbitrary $f(r)$.  Provided that only asymptotically $f(r) \ll r^2$, the spatial metric, at least at the time we are specifying the initial data, is asymptotically Lifshitz ($W(r) = r^2 + \ldots$, where the omitted terms are subleading).   This is significantly more freedom than one might have guessed and in particular, the non-normalizable piece of the electric field $f_0(r)$ has not yet been fixed.   However, given generic initial data the spacetime solution will generically evolve as a function of time and one might worry that a generic $f_0(r)$ could result in time dependence in the non-normalizable parts of the metric.  Hence we wish to check that the time evolution of the above initial data does not produce time dependence at an order much larger than the normalizable level.   Asking for the full time evolution is a question that can be addressed generically only numerically, but it is straightforward to calculate the initial acceleration of the spatial metric and this will be sufficient for our purposes.

Given the above Hamiltonian density $\mathcal{H}$ (\ref{Hform}) it is straightforward to find the field equations for the metric
\begin{equation}
\dot{h}_{a b} = \frac{\delta \mathcal{H}}{\delta \pi^{a b}} \qquad 
\dot{\pi}^{a b} = - \frac{\delta \mathcal{H}}{\delta h_{a b}}
\end{equation}
Note we are free to consider a time evolution vector consistent with the simplest Lifshitz spacetimes, namely that the shift vanishes and the lapse is only a function of $r$.   Given these assumptions and the fact that there is no initial gravitational momentum or magnetic field (although, of course, evolution generically produces both) a bit of algebra shows
\begin{eqnarray}
\ddot{h}_{a b}(0) &=& 2 N^2 \Big[ \frac{D_a D_b(N)}{N} - R^{(3)}_{a b} + h_{a b} \frac{R^{(3)}}{4} + \frac{h_{a b}}{4} \Big(2 \Lambda - \frac{m_0^2}{2} \phi^2\Big) \nonumber \\
 &-& \frac{\pi_a \pi_b}{2 \kappa^2 h} + \frac{h_{a b}}{8 \kappa^2 h} \pi^c \pi_c \Big]
\end{eqnarray}
Then plugging in the above metric (\ref{met3}) and field content (\ref{piexp}, \ref{phiconst}) the acceleration in the transverse directions is
\begin{eqnarray} \label{acceltran}
\ddot{h}_{i j}(0) &=& 2 N^2 k_{i j} \Big[ W \frac{r \partial_r N}{N} + \frac{W}{2} - \frac{R^{(0)}}{4} + \frac{\Lambda l^2 r^2}{2} - \frac{m_0^2 q^2 W}{8} \Big(1+\frac{f'}{2 r}\Big)^2 \nonumber \\
&+& \frac{q^2 z^2}{8 l^2 r^2} (r^2+f)^2\Big]
\end{eqnarray}
and in the radial direction by
\begin{eqnarray} \label{accelrad}
\ddot{h}_{r r}(0) &=& \frac{2 N^2}{W} \Big[W \frac{\partial_r^2 N}{N} + \frac{W'}{2} \frac{\partial_r N}{N} + \frac{W'}{2 r} - \frac{W}{2 r^2} + \frac{\Lambda l^2}{2} + \frac{R^{(0)}}{4 r^2} \nonumber\\
&-&\frac{m_0^2 q^2}{8 r^2} W \Big(1 +\frac{f'}{2 r}\Big)^2 - \frac{3 q^2 z^2}{8 l^2} \Big(1 + \frac{f}{r^2}\Big)^2 \Big] \nonumber \\
\end{eqnarray}
Choosing $N(r)$ such that we maintain explicit spherical/planar/hyperbolic symmetry (i.e. $\ddot{h}_{i j}(0) = 0$) and using the scalar constraint (\ref{constsol}) to determine $W'(r)$ one then finds
\begin{eqnarray} \label{raccelresult}
\ddot{h}_{r r}(0) &=& \frac{4 (z-1)}{r^2} \frac{N^2}{W} \Big(1 + \frac{f'}{2r}\Big) \Bigg[ r^2 + \frac{R^{(0)}}{4} - W \Big(1 + \frac{3 f'}{4 r} - \frac{f''}{4} \Big) \nonumber \\ 
&+& \frac{r f'}{2} \Big(1 + \frac{z}{2} + \frac{R^{(0)}}{4 r^2} \Big) - \frac{z^2 f}{2} - \frac{ z (z - 1) f^2}{4 r^2} \Big(1 + \frac{r f'}{f} + \frac{f'}{2 r} \Big) \Bigg] \nonumber \\
\end{eqnarray}
Since for these spacetimes
\be
h_{r r} = \frac{l^2}{W(r)} 
\ee
where asymptotically $W(r) \rightarrow r^2$, to maintain Lifshitz boundary conditions at a minimum we must require that $\ddot{h}_{r r}(0)$ falls off faster asymptotically than $r^{-2}$.  More generically, recalling that the normalizable part of $W(r)$, which again we recall reflects regularity in the interior and the behavior of $f(r)$ throughout the spacetime, is of order $r^{-z}$ asymptotically, then if the time evolution of the initial data does not produce time dependence beyond normalizable order it must be true that asymptotically
\be
\ddot{h}_{r r}(0) \sim r^{-z-4}
\ee
or smaller.

On the other hand, since asymptotically $N \sim r^z$ and $W \sim r^2$
\be
\frac{N^2}{r^2 W} \sim r^{2 z - 4}
\ee
then the term in brackets in (\ref{raccelresult}) must falloff faster than $r^{2- 2 z}$ to avoid breaking Lifshitz boundary conditions and falloff at least as fast as $r^{-3 z}$ if time evolution does not produce time dependence beyond the normalizable order.   However, the normalizable component of $W(r)$ at order $r^{-z}$, unless it is exactly canceled by a normalizable perturbation in $f(r)$, breaks the first condition if $z \geq 2$ and the second for all $z > 1$.   That is, if one insists upon regularity in the interior, unless one finely tunes the behavior of $f(r)$ in the interior such that solving the scalar constraint (\ref{constsol}) the resulting term in $W(r)$ exactly cancels the term at order $r^{-z}$ from $f(r)$, as well as any non-linear correction from terms falling off more slowly, the time evolution of the initial data will violate the Lifshitz boundary conditions at leading order if $z \geq 2$ and produce time dependence in the non-normalizable parts of the metric for all $z > 1$.   Finally note in the case $z = 1$, where one recovers empty AdS with no massive vector field, $\ddot{h}_{r r}(0)$ (\ref{raccelresult}) vanishes identically. This simply reflects Birkhoff's theorem or more intuitively the fact that with this much symmetry and no matter field there are no local degrees of freedom left in the spacetime.  

The reader might be concerned that the above results are contaminated by some subtlety involving orders of limits for large $r$ and small times.  The most direct way to check this concern is to do an expansion at large $r$ assuming Lifshitz asymptotics at all times and look for a similar constraint.  This is the subject of the next section.

\subsection{Asymptotic expansion}

Now we wish to consider in some level of detail the behavior of the metric and fields near infinity.  We will need the analog of the Fefferman-Graham expansion \cite{FeffermanGraham}  and hence go somewhat beyond  beyond the results in linearized perturbation theory previously obtained (\cite{RossHologLifshitz}, \cite{Peet}). For the sake of simplicitly we will confine our attention, as before, to the case of transverse symmetry (i.e. planar, spherical, or hyperbolic symmetry depending on the sign of $R^{(0)}$).  The asymptotic expressions are substantially simpler in Gaussian normal gauge where we choose $g_{r r} = r^{-2}$ and $g_{r a} = 0$ for $a \neq r$.  This differs from the previous gauge used in solving the initial data problem (where this gauge choice would transform the scalar constraint (\ref{scalconst}) into a non-linear second order differential equation) but it is straightforward to perform a diffeomorphism to compare the results.  Given the assumption of transverse symmetry, the metric will be of the form
\be
ds^2 = l^2\Big[ - r^{2 z} t_0(r,t) dt^2 + \frac{dr^2}{r^2} + r^2 a_0(r,t) k_{i j} (y) dy^i dy^j \Big]
\ee
and the massive vector field of the form
\be
A = q \, r^z b_0(r,t) dt + l \, r^{-z-1}  \alpha_r (r,t) dr
\ee
Then the Einstein and field equations yield
\begin{eqnarray}
\alpha_r &=& \frac{\dot{a}_0}{\sqrt{2 z (z - 1)} a_0 b_0} \Big[z - 1 + \frac{r a_0'}{2 a_0} + \frac{r t_0'}{2 t_0} \Big] - \frac{r \dot{a}_0'}{\sqrt{2 z (z - 1)} a_0 b_0} \, \, \, \quad \\
r^2 a_0'' &=& \frac{R^{(0)}}{2 r^2} - (z^2+z - 2) \frac{a_0}{2 t_0} ( b_0^2 - t_0) - 4 r a_0'\Big(1 - \frac{r a_0'}{16 a_0} \Big) \nonumber \\
&-& (z-1) r b_0 b_0' \frac{a_0}{t_0} \Big( 1 + \frac{r b_0'}{2 z b_0} \Big) + \mathcal{O}(r^{-2 z} \dot{\alpha}_r, r^{-2 z} {\dot{a}_0}^2)
\end{eqnarray}
\begin{eqnarray}
r t_0' &=& \frac{1}{2 z a_0 ( 2 a_0 + r a_0')} \Bigg[ -2 z (z-1)(z-2) a_0^2 (b_0^2 - t_0) + 2 z a_0 t_0 \frac{R^{(0)}}{r^2} \nonumber \\
&-& 4 z (z + 1) t_0 a_0 r a_0' \Big( 1 + \frac{r a_0'}{4 (1+z) a_0} \Big) -  4 z (z-1) a_0^2 r b_0 b_0' \Big( 1 + \frac{r b_0'}{2 z b_0} \Big)  \nonumber \\
&+& \mathcal{O}(r^{-2 z} \dot{\alpha}_r, r^{-2 z} \ddot{a}_0, r^{-2 z} {\dot{a}_0}^2, r^{-2 z} \dot{t}_0 \dot{a}_0)\Bigg]  \\
r^2 b_0'' &=& - (3+z) \Big[ 1 + \frac{ r a_0'}{(3+z) a_0} - \frac{r t_0'}{2 (3+z) t_0} \Big] r b_0'   \nonumber \\
&-& z b_0 \frac{r a_0'}{a_0} +z b_0 \frac{r t_0'}{2 t_0} + \mathcal{O}(r^{-2 z} \dot{\alpha}_r, r^{1 -2 z} \dot{a}_0' ) \\
r^2 t_0'' &=& \frac{3}{2} (z^2 + z - 2) (b_0^2 - t_0) - \frac{t_0 R^{(0)}}{2 a_0 r^2} -(z - 1) \frac{r t_0 a_0'}{a_0} \Big(1 - \frac{r a_0'}{4 (z-1) a_0} \Big)\nonumber\\
&+& 3 (z-1) r b_0 b_0' \Big(1 + \frac{r b_0'}{2 z b_0} \Big) - 2 (z+1) r t_0' \Big(1+\frac{r a_0'}{4 (z+1) a_0} - \frac{r t_0'}{4 (z+1) t_0} \Big) \nonumber \\
&+& \mathcal{O}(r^{-2 z} \ddot{\alpha}_r, r^{-2 z} \ddot{a}_0, r^{-2 z} {\dot{a}_0}^2, r^{-2 z} \dot{a}_0 \dot{t}_0) \\
r \dot{b}_0' &=& \frac{\dot{a}_0}{a_0} \Big[ \frac{z}{b_0} (t_0 - b_0^2) - r b_0' +\frac{z}{2 (z-1)}\frac{t_0 r a_0'}{a_0 b_0} + \frac{z}{2 (z-1)} \frac{r t_0'}{b_0} \Big] \nonumber \\
&+& \frac{z b_0}{2 t_0} \dot{t}_0 \Big( 1 + \frac{r b_0'}{z b_0} \Big) - \frac{z t_0 r \dot{a}_0'}{(z-1) a_0 b_0} - z \dot{b}_0 \nonumber \\
&+& \mathcal{O}(r^{-2 z} \ddot{\alpha}_r, r^{-2 z} \dot{\alpha}_r \dot{a}_0, r^{-2 z} \dot{\alpha}_r \dot{t}_0)
\end{eqnarray}
where for all the above functions $f_i(r,t)$, $f_i' = \partial_r f_i$ and $\dot{f}_i = \partial_t f_i$.  Note we further assume that asymptotically derivatives of the above functions falloff, namely
\be \label{derivassumpt}
r f_i'(r,t) \lll f_i(r,t)
\ee
Note that once one requires $a_0$, $b_0$, and $t_0$ all approach unity asymptotically,  (3.60)-(3.62) imply the solution is specified up to, at most, two time dependent functions.   In fact (3.64) restricts these functions and, as we will see below, in each case the spacetime is specified by one constant and one time dependent function.  
Solving the above equations to leading order one finds for $z \neq 2$,
\begin{eqnarray} \label{expangen}
b_0(r,t) &\approx & 1+\frac{ (z - 4) R^{(0)}}{8 (z^2 - 2 z + 2) r^2} + K_0 r^{-z-2} + C_1(t) r^{n_1} + C_2(t) r^{n_2}  \nonumber \\
t_0(r,t) &\approx &1 - \frac{(z -2) R^{(0)}}{4 (z^2 -2 z + 2) r^2} - \frac{4 (z -1) K_0}{z^2 + z + 4} r^{-2-z} + \nonumber \\
&+& \frac{(10 - z + 3 \gamma) C_1(t)}{5 z} r^{n_1}+\frac{(10 - z - 3 \gamma) C_2(t)}{5 z} r^{n_2} \nonumber \\
a_0(r,t) &\approx& 1-\frac{R^{(0)}}{4 (z^2 - 2 z + 2) r^2} - \frac{2 (z - 1) K_0}{z^2 + z + 4} r^{-z-2} \nonumber \\
&-& \frac{(3 z + \gamma) C_1(t)}{5 z} r^{n_1} + \frac{(-3 z + \gamma) C_2(t)}{5 z} r^{n_2}
 \end{eqnarray}
 where $K_0$ is a constant and
 \begin{eqnarray}
 \gamma &=& \sqrt{ 9 z^2 - 20 z + 20} \nonumber \\
 n_1 &=& -\frac{z}{2} - 1 + \frac{\gamma}{2} \nonumber \\
 n_2 &=& -\frac{z}{2} - 1 - \frac{\gamma}{2}
 \end{eqnarray}
 and for $z = 2$
 \begin{eqnarray} \label{expan2}
 b_0(r,t) &\approx &1+ \frac{R^{(0)}}{8 r^2} +  \frac{K_1 \log(r)}{r^4} + \frac{C_2(t)}{r^4} \nonumber \\
 t_0(r,t) &\approx & 1 -\frac{2 K_1\log(r)}{5 r^4} + \frac{3 K_1 - 2 C_2(t)}{5 r^4} \nonumber \\
 a_0(r,t) &\approx& 1-\frac{R^{(0)}}{8 r^2} - \frac{K_1 \log(r)}{5 r^4} - \frac{K_1 + C_2(t)}{5 r^4}
 \end{eqnarray}
 where $K_1$ is another constant.  Note the above equations have to be interpreted with a bit of care--it is not generically true that the above equations specify the asymptotics to the order of the smallest power given.  However, the pieces with free constants, provided they meet the desired boundary conditions, provide as much freedom as allowed by the generic solution, as argued above, and will be part of the fully non-linear solution.  Hence, once one adds the appropriate non-linear corrections to the desired order (and those may always be found perturbatively, order by order, using the above equations) one has the generic asympotics.
 
 Note for $z > 2$, $n_1 > 0$, resulting in a term violating the desired boundary conditions, and we must take $C_1(t) = 0$. Regarding the function $C_2(t)$, it is easy to check for $z > 2$, $n_2 < -z-2$ and so one encounters no similar difficulties.  For $z < 2$, $-0.15 \lessapprox n_1 < 0$ so then naively one might allow a nonzero $C_1(t)$ but one can check these terms will not result in a finite Hamiltonian, even if one requires the perturbations to satisfy the equations of motion asymptotically\footnote{To verify this one needs to work out the above equations to the next order, but since the magnitude of $n_1$ is sufficiently small that even if $R^{(0)} \neq 0$ second order perturbation theory suffices to find the appropriate terms.} and so must be excluded.  On the other hand, while $n_2>-z-2$ for $1 < z < 2$ and hence $C_2(t)$ would naively correspond to a non-normalizable perturbation.  However, if one requires the leading order asymptotic perturbations to satisfy the equation of motion, as in the above expansion, one does obtain a finite Hamiltonian.   Likewise for $z = 2$ if one allowed generic logarithmic perturbations of the type in the above expansion one would obtain a divergent Hamiltonian but if one insists upon precisely the above asymptotic expansion the Hamiltonian will be finite even if $K_1 \neq 0$.  This leaves us with the solution being specified, in each case, by one constant and one time dependent function (as well as the boundary condition $R^{(0)}$).   
 
 The key observation we wish to make is that the above expansions fix the normalizable part of the metric in terms of the normalizable part of the vector field.    We emphasize that we use the term normalizable here to mean those pieces which may be varied without producing a divergent Hamiltonian; we do not wish to make any assertions regarding the somewhat delicate problem in Lifshitz spacetimes of determining the appropriate bulk quantities corresponding to expectation values of operators in the proposed dual field theory \cite{RossHologLifshitz}.  To compare with the previous initial data problem one may perform a diffeomorphism to obtain manifestly spherical/planar/hyperbolic symmetric coordinates via defining a new radial coordinate $\rho = r \sqrt{a_0}$ and a new time coordinate $\tau$ to ensure $g_{\tau \rho} = 0$.  The resulting spatial metric (i.e. at fixed $\tau$) is
 \be
 ds^2 = l^2\Big[ \frac{d \rho^2}{W(\rho)} + \rho^2 k_{i j} dy^i dy^j \Big]
 \ee
 where
 \be \label{Wrho}
 W(\rho) = \rho^2 \Big( 1 + \frac{r \partial_r a_0}{2 a_0} \Big)^2 \Big[1 - \Big(\frac{\partial_t \rho}{\partial_r \rho} \Big)^2 \frac{r^{-2-2z}}{t_0} \Big]
 \ee
 and $r$ is determined in terms of $\rho$ by inverting
 \be
 \rho = r \sqrt{a_0(r,t)}
 \ee
 While one generically can not perform this inversion analytically, we will only be concerned with expansions asymptotically and in particular only need obtain $r = \rho(1+ \ldots)$ where the series is worked out to order $\rho^{-z}$.  In terms of initial data, as we have emphasized before, the coefficient of order $\rho^{-z}$ in $W(\rho)$ is determined by regularity in the deep interior and the behavior of the massive vector field throughout the bulk, not just near infinity.   For generic regular initial data the coefficient of $W(\rho)$ at order $\rho^{-z}$ is independent of the coefficient of $f(\rho)$ (or equivalently $\pi^{\rho}$) at order $\rho^{-z}$; the asymptotic behavior of the vector field is independent of its behavior in the deep interior.  Examining the expansions (\ref{expangen}) and (\ref{expan2}) one can then quickly show the term in square brackets in (\ref{Wrho}) is unity to this order and the only terms which contribute to $W$ at this order (i.e. the normaliable part) are coefficients from $a_0$ of order $r^{-z-2}$ (except in the case $z=2$ where the coefficient of the logarithmic term also enters), up to functions only of $R^{(0)}$.  Note these coefficients ($C_2$ and $K_1$ for the case $z = 2$ and $K_0$ if $z \neq 2$) are all given  in terms of the asymptotic behavior of the vector field. Of course, to make the direct comparison with initial data one should compute the leading order $\pi^\rho$ given the above asymptotics; straightforward (albeit slightly tedious) algebra shows, as one might expect, to normalizable order there are no degrees of freedom beyond those in $b_0$.

   Lest one be concerned that the above argument on the asymptotic expansion misses non-linear corrections in computing the functions $(a_0(r), t_0(r), b_0(r))$ to order $r^{-z-2}$,  we note it is possible to check this term at the fully non-linear level by using (3.60) and (3.61) directly.  For $z > 2$, the component of each of the functions at order $r^{-z-2}$ is sourced only by smaller order terms (i.e. those proportional to $R^{(0)}$) and the component of the other functions at order $r^{-z-2}$.  A few lines of algebra shows the coefficients of each of the functions of $r^{-z-2}$ are all proportional to one another (where the constant of proportionality depends only on $z$), up to possible contributions due to $R^{(0)}$.  For $z < 2$, $n_2 > -z-2$ but $2 n_2 < -z - 2$ and $n_2 - 2 < -z - 2$ and so the non-linear corrections in (3.60) and (3.61) do not change the fact that the coefficients of each of the functions of $r^{-z-2}$ are proportional to one another (again with the proportionality determined only in terms of  $z$) and the coefficients of each of the functions of $r^{-z-2}$ are all determined in terms of a single constant  (besides, of course, $R^{(0)}$).  For $z= 2$, $K_1$ can also enter into the relevant term but $K_1$ just parametrizes the asymptotic falloff of the vector field and again the metric at normalizable order is determined entirely in terms of the vector field asymptotics.  
   
   Finally we note that due to the constrained asymptotics discussed above, the mass of any asymptotically Lifshitz solution is determined entirely by the asymptotics of the vector field and independent of the behavior of the vector field in the interior of the spacetime.  In particular, assuming the above asymptotic expansion and computing directly the value of the on-shell Hamiltonian for $z \neq 2$
   \be
   E = -\frac{4 (z - 1) (z^2 - 4) \kappa l^2 \Omega_k}{z (z^2+z + 4)} K_0
   \ee
    where $\Omega_k = \int \sqrt{k^{(0)}}$ and if $z = 2$
  \be
  E = -\frac{8 \kappa l^2 \Omega_k}{5} K_1
  \ee
  if one regards $K_1$ as parametrizing a normalizable perturbation and $E = 0$ if one does not.   Note the results for $z = 2$ are independent of $C_2(t)$, which naively would determine the mass (as can be seen directly in the Hamiltonian or by noting in the previous coordinate system this terms sources a term of order $\rho^{-2}$ in $W(\rho)$), and so one does obtain a conserved energy despite the apparent time dependence at this order.\footnote{In the usual AdS case, if one does not insist upon this much symmetry (i.e. allows local gravitational degrees of freedom) there is a full tensor at the corresponding order and only its trace is time independent and enters into the mass.}   In particular, the exact and numerical black holes constructed in the $z=2$ case in \cite{Mann,Peet, Dan} have set $K_1 = 0$ (i.e. forbade asymptotic logarithmic terms) and hence are exactly massless, despite having a wide range of sizes.  The fact that the mass is constrained and independent of any perturbation suggests quite strongly the above statement that generic perturbations are not allowed is weaker than need be and in fact only exactly stationary solutions respect both Lifshitz asymptotics and regularity.
 
 \section{Closing Remarks}
 
We have pointed out that flat-section Lifshitz spacetimes with the standard behavior in the deep interior suffer from a naked singularity and further that singularity can not be resolved by either $\alpha'$ or loop effects in string theory.   Furthermore, the other obvious candidates for a ground state of flat-section asymptotically Lifshitz spacetimes, at least with the obvious symmetries, that are regular in the interior are not compatible with the Einstein and field equations.   More generically, we attempted to study the energetic stability of such spacetimes and ran into the surprising fact that Lifshitz asymptotics plus regularity in the interior of the spacetime places strong constraints on smooth perturbations in the bulk spacetime, not just asymptotically.  In particular, the mass of asymptotically Lifshitz solutions with transverse (i.e. spherical/planar/hyperbolic) symmetry is entirely determined in terms of the asymptotic behavior of the massive vector field alone and independent of any perturbation in the deep interior.   This suggests one should be able to prove no perturbations beyond the static solutions are allowed by Lifshitz asymptotics, at least if one assumes transverse symmetry, but even the above results indicate that the desired AdS/CFT behavior--where one fixes asymptotic boundary conditions but allows generic, and indeed even highly quantum, behavior in the interior--fails for Lifshitz asymptotics.

In the case of usual anti de Sitter space, the two notions of regular pertubations, one based on Hamiltonian finiteness \cite{Henneaux:1984xu, HIM} and one based on the asymptotic expansion assuming an asymptotically AdS spacetimes at all times (i.e. Fefferman-Graham \cite{FeffermanGraham}) match in the sense the asymptotic expansion leaves an undetermined tensor at the same order that determines the mass (via assuming regularity in the bulk and solving the constraints throughout the deep interior).\footnote{To the best of our knowledge the more detailed examination of the matching of interior regularity and asymptotic conditions has only been investigated in detail in four dimensions \cite{Friedrich}, although it seems reasonably clear there are no problems with standard AdS asymptotics in any number of dimensions.}   For the case of Lifshitz asymptotics, we have pointed out these conditions do not generically match.  This suggests anti-de Sitter asymptotics are rather more delicate than has been previously thought and other deformations of these asymptotics, in particular those that modify the leading order metric, deserve inquiries along the same lines.

Finally, one is left as to the question of status of Lifshitz spacetimes in string theory.   The most concrete embeddings of the above actions into a suitable supergravity solution \cite{Gauntlett08, Gauntlettetal, Gregory2010} do not present any contradiction to the above; if the solution is pathological in the four dimensional sense it will also be so, presumably, in the higher dimensional one but this has nothing to do with finding a consistent truncation.  There are by now a number of somewhat more indirect arguments that Lifshitz type solutions should be included in string theory \cite{Gubser,Joeetal}.   Hence it becomes important to understand if there are possible loopholes in these arguments, at least as applied to a full Lifshitz spacetime, or whether whether the pathologies we have described are necessarily encountered in certain string theory configurations.   Even with spacetimes which have the conventional asymptotics but flow in the interior to a Lifshitz solution, the above suggests one should check for instabilities in any region with Lifshitz scaling and singularities like those above in the case of flat slicing solutions.

\section*{Acknowledgements}
It is a pleasure to thank J. Gauntlett, R. Myers, and S. Ross for useful discussions.  This work was supported in part by the Natural Sciences and Engineering Research Council of Canada.   Research at Perimeter Institute is supported by the Government of Canada through Industry Canada and by the Province of Ontario through the Ministry of Research \& Innovation.


\begin{thebibliography}{99}

\bibitem{AdSCFT} J. M. Maldacena, ``The large N limit of superconformal field theories and supergravity'', Adv. Theor. Math. Phys. {\bf 2}, 231 (1998) [Int. J. Theor. Phys. {\bf 38}, 1113 (1999)]; S. S. Gubser, I. R. Klebanov and A. M. Polyakov, ``
Gauge theory correlators from non-critical string theory'', Phys. Lett. {\bf B428}, 105 (1998);  E. Witten, ``Anti-de Sitter space and holography'', Adv. Theor. Math. Phys. {\bf 2}, 253 (1998).



\bibitem{Kov}  P. Kovtun, D. T. Son, and A. O. Starinets,
``Viscosity in strongly interacting quantum
field theories from black hole physics'', Phys. Rev. Lett.
\textbf{94} (2005) 111601 [hep-th/0405231].

\bibitem{Hart}  S. A. Hartnoll,``Lectures on holographic methods for
condensed matter physics'', Class. Quant. Grav. {\bf 26} (2009) 224002  [arXiv:0903.3246 [hep-th]; S. A. Hartnoll, C. P. Herzog and G. T. Horowitz, ``Building a holographic superconductor'',
Phys. Rev. Lett. \textbf{101} (2008) 031601 [arXiv:0803.3295]; S. A. Hartnoll and P. Kovtun, ``Hall conductivity from dyonic black holes'',
Phys. Rev. D \textbf{76} (2007) 066001 [arXiv:0704.1160]; C. P. Herzog, ``Lectures on holographic superfluidity and superconductivity'',
J. Phys. A \textbf{42} (2009) 343001 [arXiv:0904.1975].

\bibitem{Faul}  T. Faulkner, H. Liu, J. McGreevy and D. Vegh,``Emergent quantum criticality, Fermi surfaces and AdS2'',
arXiv:0907.2694.

\bibitem{Mc}  J. McGreevy, ``Holographic duality with a view toward many-body physics'', arXiv:0909.0518.

\bibitem{dSCFT} A. Strominger, ``The dS / CFT correspondence''
 JHEP {\bf 0110} 034 (2001);
V. Balasubramanian, J. de Boer and D. Minic,
``Mass, entropy and holography in asymptotically de Sitter spaces'',
Phys.Rev. {\bf D65}  123508 (2002);
A.M. Ghezelbash and R.B. Mann,
``Action, mass and entropy of Schwarzschild-de Sitter black holes and the de Sitter / CFT correspondence'', JHEP {\bf 0201} 005 (2002).
\bibitem{MannMarolf}   R.~B.~Mann and D.~Marolf,
  ``Holographic renormalization of asymptotically flat spacetimes'',
  Class.\ Quant.\ Grav.\  {\bf 23}, 2927 (2006);
  D.~Marolf, ``Asymptotic flatness, little string theory, and holography'',
JHEP {\bf 0703} 122 (2007); R.~B.~Mann, D.~Marolf, R. McNees and  A. Virmani
``On the Stress Tensor for Asymptotically Flat Gravity'', Class. Quant. Grav. {\bf 25} 225019 (2008).

\bibitem{Kachru}  S. Kachru, X. Liu and M. Mulligan, ``Gravity duals of Lifshitz-like fixed points'', Phys. Rev. D \textbf{78}
(2008) 106005[arXiv:0808.1725].

\bibitem{Koroteev} P. Koroteev and M. Libanov, ``On Existence of Self-Tuning Solutions in Static Braneworlds without Singularities'', JHEP {\bf 0802} (2008) 104 [arXiv:0712.1136].


\bibitem{Tay}  M. Taylor, ``Non-relativistic holography'', arXiv:0812:0530.

\bibitem{Mann}  R. B. Mann, ``Lifshitz topological black holes'', JHEP \textbf{0906} (2009) 075 [arXiv:0905.1136].

\bibitem{Peet}  G. Bertoldi, B. A. Burrington and A. Peet, ``Black holes in
asymptotically Lifshitz spacetimes with arbitrary critical exponent'', Phys. Rev. D {\bf 80} (2009) 126003
[arXiv:0905.3183].

\bibitem{Bal}  K. Balasubramanian and J. McGreevy, ``An analytic Lifshitz
black hole'', Phys. Rev. D {\bf 80} (2009) 104039  [arXiv:0909.0263]].

\bibitem{Dan} U. H. Danielsson, L. Thorlacius,
``Black holes in asymptotically Lifshitz spacetime'',  JHEP \textbf{0903} (2009) 070 [arXiv:0812.5088].

\bibitem{Horava} P. Ho\v{r}ava, ``Quantum Gravity at a Lifshitz Point'', Phys. Rev. D 79, 084008 (2009).

\bibitem{Gauntlett08}
  A.~Donos and J.~P.~Gauntlett, ``Lifshitz Solutions of D=10 and D=11 supergravity,''
  arXiv:1008.2062 [hep-th].


   \bibitem{topcensor} J. L. Friedman, K. Schleich, and D. W. Witt, ``Topological Censorship,'' Phys. Rev. Lett. {\bf 71} (1993), 1486  [arXiv:gr-qc/9305017]; G. J. Galloway, K. Schleich, D. M. Witt, and E. Woolgar, ``The AdS/CFT Correspondence Conjecture and Topological Censorship,'' Phys. Lett. {\bf B505} 255-262(2001)  [arXiv:hep-th/9912119]

\bibitem{Magoo}
  O.~Aharony, S.~S.~Gubser, J.~M.~Maldacena, H.~Ooguri and Y.~Oz,
  `Large N field theories, string theory and gravity,''
  Phys.\ Rept.\  {\bf 323}, 183 (2000)
  [arXiv:hep-th/9905111].



\bibitem{posenergy}
  E.~Witten,
``A Simple Proof Of The Positive Energy Theorem,''
  Commun.\ Math.\ Phys.\  {\bf 80}, 381 (1981) ; 
 G.~W.~Gibbons, C.~M.~Hull and N.~P.~Warner, ``The Stability of Gauged Supergravity,'' Nucl. Physi. B {\bf 218} (1983) 173; W. Boucher, ``Positive Energy Without Supersymmetry,'' Nucl. Physi. B {\bf 242} (1984) 282; P.~K.~Townsend, ``Positive Energy and the Scalar Potential in Higher Dimensional (Super)Gravity Theories,'' Phys. Lett. B {\bf 148} (1984) 55
 
  
       \bibitem{Wittenbubble}
 E.~Witten,
  ``Instability Of The Kaluza-Klein Vacuum,''
  Nucl.\ Phys.\ B {\bf 195} (1982) 481; 
D.~Brill and H.~Pfister, ``States of negative total energy in Kaluza-Klein theory,'' Phys. Lett. B {\bf 228}, 359 (1989); D.~Brill and G.~T.~Horowitz, ``Negative energy in string theory,'' ' Phys. Lett. B {\bf 262}, 437 (1991).

\bibitem{CopseyMannHOP}
K. Copsey and R. B. Mann, ``States of Negative Energy and $AdS_5 \times S_5/Z_k$'', JHEP {\bf 0805} 069 arXiv:0803.3801.




\bibitem{Horosteif} G. Horowitz and A. Steif, ``Space-Time Singularities in String Theory,'' Phys. Rev. Lett. {\bf 64} (1990) 260.

\bibitem{Gregory2010}
  R.~Gregory, S.~L.~Parameswaran, G.~Tasinato and I.~Zavala,
  ``Lifshitz solutions in supergravity and string theory,''
  JHEP {\bf 1012}, 047 (2010)
  [arXiv:1009.3445 [hep-th]].




\bibitem{HorowitzRoss} G. T. Horowitz and S. F. Ross, ``Naked Black Holes,'' Phys. Rev. {\bf D56} (1997) 2180-2187.  arXiv: hep-th/9704058

\bibitem{Gubser:2000nd}
  S.~S.~Gubser,
  `Curvature singularities: The good, the bad, and the naked,''
  Adv.\ Theor.\ Math.\ Phys.\  {\bf 4}, 679 (2000)
  [arXiv:hep-th/0002160].


\bibitem{HorowitzMyers} G. T. Horowitz and R. Myers ``The Value of Singularities, '' Gen.Rel.Grav.{\bf 27} 915-919 (1995). arXiv: gr-qc/9503062 

\bibitem{RossHologLifshitz} S. F. Ross and O. Saremi, ``Holographic stress tensor for non-relativistic theories,'' arXiv:  0907.1846


\bibitem{Deh3} M. H. Dehghani and R. B. Mann, ``Thermodynamics of Lovelock-Lifshitz Black Branes'',
Phys. Rev. {\bf D}  (to be published) arXiv:1006.3510 [hep-th].


\bibitem{Arnowitt:1962hi}
  R.~L.~Arnowitt, S.~Deser and C.~W.~Misner,
 `The dynamics of general relativity,'' in ÒGravitation: An Introduction to Current
ResearchÓ (L. Witten, Ed.), John Wiley and Sons, New York, 1962.
  arXiv:gr-qc/0405109

\bibitem{Henneaux:1984xu}
  M.~Henneaux and C.~Teitelboim,
``Hamiltonian Treatment Of Asymptotically Anti-De Sitter Spaces,''
  Phys.\ Lett.\  B {\bf 142}, 355 (1984).
  
  \bibitem{topBH}
J.P.~Lemos,
`Cylindrical black hole in general relativity,"
Phys. Lett. {\bf B353}, 46 (1994)
gr-qc/9404041; J.P.~Lemos, ``Two-dimensional black holes and planar general relativity,"
Class. Quant. Grav. {\bf 12}, 1081 (1995)
gr-qc/9407024; S.~Aminneborg, I.~Bengtsson, S.~Holst and P.~Peld{\'a}n,``Making anti-de Sitter black holes,"
Class. Quant. Grav. {\bf 13}, 2707 (1996)
gr-qc/9604005; R.B.~Mann, ``Pair production of topological anti-de Sitter black holes,"
Class. Quant. Grav. {\bf 14}, L109 (1997)
gr-qc/9607071; D.~Brill, J.~Louko and P.~Peld{\'a}n, 
Phys. Rev. {\bf D56}, 3600 (1997)
gr-qc/9705012; L.~Vanzo, ``Black holes with unusual topology,"
Phys. Rev. {\bf D56}, 6475 (1997)
gr-qc/9705004; R.B.~Mann, ``Topological Black Holes: Outside Looking In,"
gr-qc/9709039.



\bibitem{Boer} J. de Boera, M. Kulaxizia, and A. Parnachev, ``Holographic Lovelock Gravities and Black Holes'',
arXiv:0912.1877 [hep-th].

\bibitem{Cam} X. O. Camanho and J. D. Edelstein, ``Causality in AdS/CFT and Lovelock theory'',
arXiv:0912.1944 [hep-th].

\bibitem{Deh2} M. H. Dehghani and R. B. Mann,``Lovelock-Lifshitz Black
holes'', JHEP \textbf{1007}  (2010) 019 [arXiv:1004.4397].

\bibitem{Oliva}G. Dotti, J. Oliva, and R. Troncoso,``Exact solutions for the Einstein-Gauss-Bonnet thory in five dimensions:
Black holes, wormholes and spacetime horns'',  Phys. Rev. D \textbf{76} (2007) 064038 [arXiv:0706.1830].

\bibitem{Ay1}  E. Ayon-Beato, A. Garbarz, G. Giribet and M. Hassaine,
``Lifshitz black hole in three Dimensions'', Phys.
Rev. D \textbf{80} (2009) 104029 [arXiv:0909.1347].

\bibitem{Ay2} E. Ayon-Beato, A. Garbarz, G. Giribet and M. Hassaine,
``Analytic Lifshitz black holes in higher dimensions'', JHEP \textbf{1004} (2010) 030 [arXiv:1001.2361].

\bibitem{Cai} R. G. Cai, Y. Liu and Y. W. Sun,``A Lifshitz black hole in four dimensional $R^2$ gravity'',
JHEP \textbf{0910} (2009) 080 [arXiv:0909.2807].

\bibitem{Pang} D. W. Pang,``On charged Lifshitz black holes'', JHEP \textbf{1001} (2010) 116 [arXiv:0911.2777].

\bibitem{Myers} R. C. Myers and B. Robinson, ``Black Holes in Quasi-topological Gravity'', arXiv:1003.5357 [gr-qc].

\bibitem{Deh} M. H. Dehghani and M. Shamirzaie, ``Thermodynamics of asymptotic flat charged black holes in third order Lovelock gravity'', Phys. Rev. D {\bf 72}, 124015 (2005).
\bibitem{Iyer} V. Iyer and R. M. Wald, ``Some Properties of Noether Charge and a Proposal for Dynamical Black Hole Entropy'', Phys. Rev. D {\bf 50} (1994) 846 [gr-qc/9403028].

\bibitem{Deh4} M. H. Dehghani and R. Pourhasan, ``Thermodynamic Instability of Black Holes of Third Order Lovelock Gravity'', arXiv:0903.4260.

 \bibitem{HIM}
 S. Hollands, A. Ishibashi, and D. Marolf, ``Comparison Between Various Notions of Conserved Charges in Asymptotically AdS-Spacetimes,'' Class. Quant. Grav. 22 (2005) 2881-2920.

\bibitem{FeffermanGraham}
{C. Fefferman and C. Robin Graham, ``Conformal Invariants'', in 
{\it Elie Cartan et les Math\'ematiques d'aujourd'hui} (Ast\'erisque, 1985) 
95.}

\bibitem{Friedrich}
  H.~Friedrich, ``Einstein equations and conformal structure - Existence of anti de Sitter type space-times,''
  J.\ Geom.\ Phys.\  {\bf 17}, 125 (1995).

\bibitem{Gauntlettetal} 
  A.~Donos, J.~P.~Gauntlett, N.~Kim and O.~Varela, ``Wrapped M5-branes, consistent truncations and AdS/CMT,''
  arXiv:1009.3805 [hep-th].


\bibitem{Gubser}
  S.~S.~Gubser and A.~Nellore, ``Ground states of holographic superconductors,''
  Phys.\ Rev.\  D {\bf 80}, 105007 (2009)
  [arXiv:0908.1972 [hep-th]].
  
\bibitem{Joeetal}
  S.~A.~Hartnoll, J.~Polchinski, E.~Silverstein and D.~Tong,
  `Towards strange metallic holography,''
  JHEP {\bf 1004}, 120 (2010)
  [arXiv:0912.1061 [hep-th]].



\end{thebibliography}
\end{document}